\RequirePackage{fix-cm}
\documentclass[a4paper]{jpconf}
\usepackage{graphicx}
\usepackage{amsmath,amssymb,amsfonts}
\usepackage{hyperref}
\usepackage{tikz}
\usepackage{cite}

\begin{document}

\title{Momentum entanglement in relativistic quantum mechanics}

\author{W~Smilga}

\address{Isardamm~135~d, 82538~Geretsried, Germany}

\ead{wsmilga@compuserve.com}

\begin{abstract}
I present a new group-theoretical approach to the interaction mechanism of elementary particle 
physics. 
Within an irreducible unitary two-particle representation of the Poincar\'e group, 
the commutation relations of the Poincar\'e group require that the two-particle states 
be momentum entangled. 
As in gauge theories, momentum entanglement defines a correlation between two particles
that can be described as an interaction provided by the exchange of virtual (gauge) quanta.
The coupling constant of this interaction is uniquely determined by the structure 
of the irreducible two-particle state space. 
For two massive spin one-half particles, the coupling constant matches the empirical 
value of the electromagnetic coupling constant.
%\keywords{Poincar\'e group \and interaction \and quantum electrodynamics \and quantum gravity \and 
%conformal gravity \and gauge invariance \and coupling constant}
\end{abstract}

\section{Introduction}
\label{section_1}

It is generally accepted that the fundamental interactions of elementary 
particle physics are the result of symmetries.
In the Standard Model (of particle physics), the symmetry with respect to 
gauge trans\-formations is made responsible for the electromagnetic, weak, 
and strong interactions.
It has been suggested that also the gravitational interaction can be based 
on gauge invariance, although to date no such theory exists. 
So far, attempts to derive gauge invariance from other, more basic principles
have failed.
Therefore, within the Standard Model, gauge invariance is understood as a 
first principle.

The Standard Model implements gauge invariance by coupling (charged) fermi\-ons 
to gauge fields.
Within the perturbation algorithm, this coupling is described by the exchange
of gauge bosons between two fermions.
Looking at the structure of Feynman graphs, for example of electron scattering, it 
becomes apparent that it is not the gauge symmetry itself, but rather the 
momentum entangled structure generated by the exchange of virtual gauge bosons 
that defines the interaction, independently of what may have caused the entanglement.
Hence, a more appropriate question than that of why Nature is gauge 
invariant, is the question: why are two-fermion states momentum entangled?

In the following, I will describe a fundamental mechanism that is inherent to 
relativistic quantum theories.
It enforces, under very general conditions and independently of gauge 
invariance, momentum entangled two-particle states.
I will show that this entanglement is responsible for at least two interactions: 
the electromagnetic and a gravitational interaction.

\section{Short review of the Poincar\'e group}
\label{section_2}

As a reminder, here is an excerpt from \cite{sss}:
The Poincar\'e group PO(3,1) is a ten-parameter continuous group.
The four generators for the infinitesimal translations are (represented by) 
Hermitian operators $p_\mu$.  The six Hermitian operators $m_{\mu\nu}$ generate 
the rotations in the $x^\mu-x^\nu$ plane.
The rotations form the (homogeneous) Lorentz group SO(3,1).
The commutation rules of these operators are
\begin{eqnarray}
\left[p_\mu,p_\nu \right] &=& 0 \label{2-1a} \\
\left[m_{\mu\nu},m_{\rho\sigma} \right] &=&
-i(g_{\mu\rho} m_{\nu\sigma} - g_{\nu\rho} m_{\mu\sigma}  
+g_{\mu\sigma} m_{\rho\nu} - g_{\nu\sigma} m_{\rho\mu})  \label{2-1b} \\
\left[m_{\mu\nu},p_\sigma \right] &=& i(g_{\nu\sigma} p_\mu - g_{\mu\sigma} p_\nu) .
\label{2-1}
\end{eqnarray}
The Hermitian operators $p_\mu$ and $m_{\mu\nu}$ are identified with the observables
of 4-momentum and angular momentum, respectively.
The following scalar operators (Casimir operators)
\begin{equation}
P = p^\mu p_\mu     \label{2-2}
\end{equation}
and
\begin{equation}
W = -w^\mu w_\mu , \;\;\mbox{ with }\;\;
w_\sigma = \frac{1}{2} \epsilon_{\sigma\mu\nu\lambda} m^{\mu\nu} p^\lambda,
\label{2-3}
\end{equation}
commute with all the infinitesimal generators.
Hence, for any irreducible representation of the Poincar\'e group, they are multiples 
of the identity, and their eigenvalues can be used to classify the irreducible 
representations.
Within an irreducible representation, the states can be labelled by the eigenvalues
of a complete set of commuting operators.
For a complete set, the three spatial components of the momentum 
operator $p_\mu$ and one of the components of $w_\mu$, say $w_0$, can be chosen.
In vector notation (bold letters stand for 3-vectors):
\begin{eqnarray}
&w^0& = \mathbf{p} \cdot \mathbf{m} , \\
&\mathbf{w}& = p_0 \, \mathbf{m} - \mathbf{p} \times \mathbf{n} , \label{2-4}
\end{eqnarray}
where
\begin{eqnarray}
&\mathbf{m}& = (m_{32}, m_{13}, m_{21}) ,\\
&\mathbf{n}& = (m_{01}, m_{02}, m_{03}) .    \label{2-5}
\end{eqnarray}
Given an eigenstate of momentum, $w_0$ is proportional to the generator of the 
rotation in the plane perpendicular to the 3-momentum $\mathbf{p}$.
These rotations are the only rotations that leave $\mathbf{p}$ invariant. 
For $\mathbf{p} = 0$, the rotational axis is arbitrary.

\section{Uncorrelated versus correlated particles}
\label{section_3}

A system of two independent particles with individual 4-momenta $p_1$ and $p_2$ 
is described by a product representation of the Poincar\'e group.
The momenta $p_1$ and $p_2$ refer to two independent reference frames. 
The two-particle system is subject to changes of the frames of reference
by the operations of SO(3,1) $\otimes$ SO(3,1).
As long as the frames of reference are not correlated, I will call such a
two-particle system {\it uncorrelated}.

In contrast, if the two particles are considered in the same space--time domain,
their momenta refer to one and the same frame of reference.
Whereas in an uncorrelated system the sum of the particle momenta does not
make any sense, now the total momentum $p$ can be defined by forming the sum 
$p = p_1 + p_2$.
Furthermore, from the momenta $p_1$ and $p_2$ and the particle positions 
$x_1$ and $x_2$, an orbital angular momentum, described by the operator
\begin{equation}
m^{\mu\nu} = x^\mu q^\nu - x^\nu q^\mu, \mbox{ where }
q^\mu = p^\mu_2 - p^\mu_1 \mbox{ and }
x^\mu = x^\mu_2 - x^\mu_1 , \label{3-1}
\end{equation} 
can be constructed. 
Two particles with a common frame of reference will be referred to as
{\it correlated}. 

The total momentum of a correlated two-particle system is subject to changes of 
the frame of reference by operations of a single SO(3,1) group.
In addition, there is an internal rotational degree of freedom, similar to
the spin degree of freedom of single particles.
This degree of freedom is defined by rotations in the plane perpendicular to the 
3-momentum $\mathbf{p}$, which leave the total momentum invariant but, in general, 
change the individual particle momenta.

For correlated two-particle systems, simultaneous eigenstates of the total momentum 
and of $w_0$ can be prepared.
Now, $w_0$ refers to the internal rotational degree of freedom, possibly in combination 
with the spin of the particles. 
Under the action of SO(3,1), such an eigenstate can generate a complete basis for the state 
space of an irreducible two-particle representation of PO(3,1).
Such a representation is characterized by the eigenvalues of the Casimir operators
$P$ and $W$.

An immediate consequence of the commutation relations (\ref{2-1}) between the linear
and the angular momentum is that the individual particle momenta in the operator of 
the orbital angular momentum (\ref{3-1}) do not commute with the angular momentum 
that they create:
\begin{equation}
\left[p^\nu_k, m^{\mu\nu}\right]\;\not=\;0\;\not=
\;\left[p^\mu_k, m^{\mu\nu}\right] , \;\; k = 1,2 . \label{3-2}
\end{equation}
This means that a simultaneous eigenstate of the total momentum and of the orbital 
angular momentum cannot, at the same time, be an eigenstate of $\mathbf{p}_1$ and 
$\mathbf{p}_2$, except when the 3-momenta $\mathbf{p}_1$ and $\mathbf{p}_2$, and 
therefore also $\mathbf{p}$, are parallel.
This eigenstate must, therefore, have the structure of a momentum entangled
superposition of product states.
More precisely: since an eigenstate of a component of the angular momentum is
invariant under the rotations generated by this component, the state is a
rotationally symmetric superposition of product states with the same total momentum.
In consequence, measurements of the individual particle momenta will yield 
non-deterministic results.
Nevertheless, if the state is an eigenstate of the total momentum, the particle 
momenta will always add up to the (conserved) value of the total momentum.
Hence an unbiased experimenter will conclude that the particles exchange momentum.

\section{A complementary dynamical law}
\label{section_4}

The result of the last section can be extended to all states of an irreducible
two-particle representation of the Poincar\'e group:
Since a complete set of momentum entangled eigenstates of total momentum and angular 
momentum form a basis, the entangled structure is passed on to all states of the same
irreducible representation.
If it were possible to construct from this basis a pure product state, then, because
of Poincar\'e covariance, it would be possible to construct a complete set of such 
product states.
These states would form a basis of a subspace of the product representation that is 
restricted only by the constancy of the Casimir operator $P$, but not of $W$, in 
contradiction to the assumed irreducibility. 

Therefore, a general rule holds:\\
\begin{em}The states of an irreducible two-particle representation of the Poincar\'e 
group have the form of a momentum entangled superposition of product states.\end{em}\\
Another formulation of this rule, in the pictorial language of relativistic 
perturbation theories, e.g., quantum electrodynamics (QED), is this:\\
\begin{em}In an irreducible two-particle representation of the Poincar\'e 
group, the particles exchange virtual quanta of momentum. \end{em}

An exception to this rule is the representation where the 
3-momenta $\mathbf{p}_1$ and $\mathbf{p}_2$, and therefore also $\mathbf{p}$,
are parallel. 
In this representation, there is no entanglement and hence no exchange of virtual 
quanta.

In conjunction with the conservation laws for total momentum and angular 
momentum, this rule has the character of a \begin{em}complementary 
dynamical law\end{em}, dictated by Poincar\'e symmetry.
It defines a mechanism that forces two particles to exchange quanta of momentum 
in a controlled way.
I call this rule `a law' (in the following: `the Dynamical Law') for two reasons:
Firstly, it derives directly from the commutation relations of the Poincar\'e group
and is, therefore, fundamental for relativistic quantum mechanics, as fundamental 
as the conservation laws of energy, momentum, and angular momentum.
Secondly, it has, just as the mentioned conservation laws, far-reaching implications
for the physics of elementary particles. 

Actually, the Standard Model uses a concept of momentum exchange to describe 
three of the fundamental forces between elementary particles.
Feynman, when formulating the perturbation theory of quantum electrodynamics 
\cite{rf1,rf2,rf3}, spoke of `virtual quanta'.
Based on the postulate of gauge invariance, the exchanged quanta became 
`gauge bosons'.
Integration over the exchanged momenta is an integral part of the Feynman rules 
in momentum space (cf., e.g., \cite{sss1}).
Together with the conservation of momentum at each vertex, this integration 
generates the kind of momentum entanglement that is demanded by the Dynamical Law.

As just shown, there is no need to explicitly postulate the presence of gauge 
bosons: they are there not for reasons of gauge invariance, but because the 
structure of the Poincar\'e group requires their existence.
However, in contrast to the common understanding of QED, these gauge boson are 
not the quanta of an independent field.
They rather stand for a field that describes the intrinsic structure of 
many-particle states as determined by Poincar\'e symmetry. 
Therefore, this field may be referred to as an {\it intrinsic gauge field}.

\section{Irreducible two-particle states}
\label{section_5}

The parameter space of an irreducible two-particle representation derives from
the two-particle mass-shell relation
\begin{equation}
(p_1 + p_2)^2 = M^2  ,   \label{5-1}
\end{equation}
which defines a seven-dimensional surface in an eight-dimensional space--time continuum.
After inserting the mass-shell relations for the individual particles, this parameter space
is described by a five-dimensional space $\Omega$ of independent momentum parameters.
$\Omega$ is embedded into the six-dimensional momentum space $\mathbb{R}^3\times\mathbb{R}^3$. 
It has spherical properties with respect to four dimensions and is unbounded in 
the radial direction.
Accordingly, $\Omega$ has a spherical infinitesimal volume element 
$d\omega(\mathbf{p}_1,\mathbf{p}_2)$. 
On the other hand, the embedding into $\mathbb{R}^3\times\mathbb{R}^3$
induces a Cartesian infinitesimal volume element $d^5\!p$ on $\Omega$.

Since the two-particle states of an irreducible representation are momentum entangled,
a two-particle state is represented by an integral
\begin{equation}
\left|\Phi\right> = \! \int\!d\omega(\mathbf{p_1},\mathbf{p_2})\; 
c_{\mathbf{p_1}\mathbf{p_2}}\,\left|\mathbf{p_1},\mathbf{p_2}\right>  
\label{5-2} 
\end{equation}
over pure product states $\left|\mathbf{p_1},\mathbf{p_2}\right>$, with state-specific
coeffients $c_{\mathbf{p_1}\mathbf{p_2}}$.
The range of integration is a finite (bounded) subspace of the full parameter space 
$\Omega$.
Since the product states are normalized to $\mathbb{R}^3\times\mathbb{R}^3$, a
normalization factor that maintains the normalization of the two-particle state is required.
It equals the square root of the volume of the integration area.
This area is essentially the spherical subspace of $\Omega$.
It makes sense to split the normalization factor into a state-independent factor $\omega$, 
directly related to the square root of the volume of the spherical subspace, and a 
state-specific factor.
In state (\ref{5-2}), the first factor is assumed to be included in the volume element 
$d\omega$, the second one in the coefficients $c_{\mathbf{p_1}\mathbf{p_2}}$.
The spherical volume element can then be replaced by the Cartesian volume element $d^5\!p$
\begin{equation}
d\omega(\mathbf{p}_1,\mathbf{p}_2) \;\;\Rightarrow\;\; \omega \, d^5\!p .  \label{5-3}
\end{equation}
Using (\ref{5-3}), a two-particle state can be written as
\begin{equation}
\left|\Phi\right> = \! \int_\Omega\! \omega \, d^5\!p\;
c_{\mathbf{p_1}\mathbf{p_2}}\,\left|\mathbf{p_1},\mathbf{p_2}\right>.
\label{5-4} 
\end{equation}

\section{Scattering experiments}
\label{section_6}

The experimental setup of a typical scattering experiment (Fig.\,1) will help to reveal 
the physical meaning of $\omega$.
An incoming plane wave (lower left) with momentum $\mathbf{p}$ passes an aperture 
and hits a target (centre).
Parts of the scattered wave pass the second aperture and are registered by 
a detector (right).
Between the apertures, an incoming particle together with a particle of the target 
form an intermediate, correlated two-particle system.
By positioning the first aperture at a distance $\mathbf{d}$ from the 
target, an orbital angular momentum $\mathbf{d} \times \mathbf{p}$ is selected.
The observed transition is represented by a hyperbolic trajectory.

%\vspace{0.2cm}
\hspace{-0.4cm}{
\begin{center}
\begin{tikzpicture}[scale=1.2]
		
\draw[rotate=45,black,very thick] (3,0.95) -- (3,0.05);
\draw[rotate=45,black,very thick] (3,-0.15) -- (3,-1.05);
\draw[rotate=45,black,very thick] (3.03,0.95) -- (3.03,0.05);
\draw[rotate=45,black,very thick] (3.03,-0.15) -- (3.03,-1.05);

\draw[rotate=0,black,very thick] (6.63,4.90) -- (6.63,4.00);
\draw[rotate=0,black,very thick] (6.63,3.80) -- (6.63,2.90);
\draw[rotate=0,black,very thick] (7.63,4.90) -- (7.63,4.00);
\draw[rotate=0,black,very thick] (7.63,3.80) -- (7.63,3.50);
\draw[rotate=0,black,very thick] (7.63,3.10) -- (7.63,2.90);
\draw[rotate=0,black,very thick] (6.60,4.90) -- (6.60,4.00);
\draw[rotate=0,black,very thick] (6.60,3.80) -- (6.60,2.90);
\draw[rotate=0,black,very thick] (7.60,4.90) -- (7.60,4.00);
\draw[rotate=0,black,very thick] (7.60,3.80) -- (7.60,3.50);
\draw[rotate=0,black,very thick] (7.60,3.10) -- (7.60,2.90);

\foreach \x in {10,...,20}
    \draw[rotate=45,black] (\x/7,0.95) -- (\x/7,-1.05);
\filldraw [green] (3.9,4.1) circle (2pt);
\foreach \x in {3,...,19}
    \draw[rotate=45, red] (3.0 + \x/7, -0.2 - \x/7) arc (-45:45:0.2+\x/5);				
\foreach \x in {0,...,9}
    \draw[red] (4.02+\x/7, 4.25+\x/7) arc (45:-135:0.2+\x/5);
\draw [black] (2.0,0.8) node [right,text width=5cm,text centered]{Incoming plane wave};

\draw [blue, thick, x=0.02cm, y=0.85cm,
declare function={
hyperbel(\t,\a,\b)=sqrt(\t*\t*\a + \b); }] 
plot [domain=-115:120, samples=50, smooth, rotate=201.5, xshift=-5.15cm, yshift=-2.6cm] (\x - 4.0,{hyperbel(\x/5,0.0031,1.0)}); 
		
\foreach \x in {33,...,43}
    \draw[blue] (\x/5,3.80) -- (\x/5,4.00);
\draw [black] (6.7,3.30)  node [right,text width=4cm,text centered]{Outgoing plane wave};

\draw [blue, thick, x=0.02cm, y=0.85cm, declare function={hyperbel(\t,\a,\b)=sqrt(\t*\t*\a + \b);}] 

plot [domain=-115:120, samples=50, smooth, rotate=201.5, xshift=-5.15cm, yshift=-2.6cm] (\x - 4.0,{hyperbel(\x/5,0.0031,1.0)}); 

\draw[black, thin] (3.9,4.1) -- (6.5,4.1);
\draw[black, thin] (3.9,4.1) -- (2.05,2.25);
\draw[rotate=45,black,thick] (3.3,0.35) -- (3.4,0.15) -- (3.5,0.35);
\draw[rotate=45,black,thick] (3.3,-0.20) -- (3.4,-0.00) -- (3.5,-0.20);
\draw[rotate=45,black,thin] (2.0,0.0) -- (4.0,0.0);
\draw [black] (2.2,2.1)   node [right,text width=0.8cm,text centered]{$\mathbf{d}$};
\draw [black] (0.6,1.2)   node [right,text width=1cm,text centered]{$\mathbf{p}$};
\draw [black] (2.2,4.20)   node [right,text width=1.5cm,text centered]{Target};
\draw [black] (8.4,3.90)   node [right,text width=2.0cm,text centered]{Detector};

\end{tikzpicture}
\end{center}
}
%\vspace{-0.2cm}
{\bf Fig.\,1} Selection of orbital angular momentum $\mathbf{d} \times \mathbf{p}$ 
\vspace{0.3cm}

The transition amplitude from an incoming two-particle state, described by a pure 
product state $\left|\mathbf{p}_1,\mathbf{p}_2\right>$, through an intermediate 
state with well-defined total momentum $\mathbf{p}$ and angular momentum $m$, 
to the outgoing product state 
$\left|\mathbf{p}_1\!-\!\mathbf{k},\mathbf{p}_2\!+\!\mathbf{k}\right>$,
is then given by
\begin{equation}
S(\mathbf{k}) = \omega^2\left<\mathbf{p}_1,\mathbf{p}_2\right|
\left.\!\mathbf{p},m\right>\!\left<\mathbf{p},m\!\right.
\left|\mathbf{p}_1\!-\!\mathbf{k},\mathbf{p}_2\!+\!\mathbf{k}\right> . \label{6-1}
\end{equation}
Here, I have placed the state-independent normalization factor $\omega^2$ of 
$\left|\mathbf{p},m\right>\!\left<\mathbf{p},m\right|$ in front of the amplitude, which 
gives $\omega^2$ the clear meaning of a {\it coupling constant}:
$\omega$ determines, as a state-independent factor, the transition amplitude from an 
incoming product state $\left<\mathbf{p}_1,\mathbf{p}_2\right|$ to the same product state 
within $\left|\mathbf{p},m\right>$, 
and as well the transition amplitude from a product state 
$\left<\mathbf{p}_1\!-\!\mathbf{k},\mathbf{p}_2\!+\!\mathbf{k}\right|$ within 
$\left<\mathbf{p},m\right|$ to the same outgoing product state.
Because the intermediate eigenstate of total and orbital angular momentum is momentum
entangled, there will be non-zero transition amplitudes for $\mathbf{k}\not=\mathbf{0}$.

Generally speaking, any restriction on the relative momenta and positions of two correlated
particles means a restriction of the available intermediate states.
It is this restriction that makes the momentum exchange visible.

\section{Coupling constant}
\label{section_7}

As indicated above, $\omega^2$ is determined by the geometry of the two-particle mass-shell 
relation (\ref{5-1}), which is a little bit tricky, but, nevertheless, well defined.
This allows calculating $\omega^2$ from the volumes of certain homogeneous domains \cite{hua}
$D^5$, $Q^5$ and $S^4$, which have been used in \cite{ws2} to describe the structure of 
the two-particle mass-shell.
An outline of the calculation can be found in the Appendix.
The result (for massive leptons) is given by the expression
\begin{equation}
8 \pi\,V(D^5)^{\frac{1}{4}} \, / \, (V(S^4) \, V(Q^5)) 
= \frac{9}{8 \pi^4} \! \left(\frac{\pi^5}{2^4 \, 5!}\right)^{1/4}  
= 1/137.03608245.                                 	\label{7-1}	         	
\end{equation}
This expression is identical to the well-known semi-empirical Wyler formula \cite{wyl}.
Its numerical value closely matches the electromagnetic fine-structure constant 
$\alpha$, whose empirical value is 1/137.035999084(51) \cite{han}.

This close match is strong evidence that the electromagnetic interaction is 
in fact the manifestation of the intrinsic gauge field (cf.\,Section 4) required by 
the Dynamical Law.

\section{The Dynamical Law in particle theories}
\label{section_8}

Since the Dynamical Law derives directly from Poincar\'e invariance, it is binding
on every relativistic quantum theory of particle physics, including the existing gauge 
theories, as well as any future theory of quantum gravity.
This, plus the fact that the concept of Poincar\'e invariance is experimentally 
extraordinarily well supported, makes the Dynamical Law a powerful instrument for the 
evaluation of existing theories and for setting up new theories.

In the following two subsections, I will analyse how two well-known relativistic 
theories comply with the Dynamical Law.

\subsection{Quantum electrodynamics}
\label{section_8.1}

Quantum electrodynamics is presently implemented with a perturbation algorithm. 
This algorithm has inconsistencies, in the form of divergent integrals 
(cf., e.g., \cite{sss2}).
The technique of renorm\-alization was developed to remove these divergences.
Since renormalization is more a recipe than a mathematical procedure, it makes the 
algorithm lack transparency.
Nevertheless, the numerical results obtained in this way match the experimental data 
extremely well.

In contrast, the Dynamical Law offers a non-perturbative way to obtain the 
electromagnetic interaction. 
This makes it well suited to shed some light on the perturbation algorithm.

The perturbation algorithm of QED constructs states with a momentum entangled 
structure, in agreement with the structure of the states of irreducible 
two-particle representations.
It also correctly normalizes the intermediate states by inserting the empirical 
value of the fine-structure constant $\alpha$ as the coupling constant.
However, instead by restricting the product state space to an irreducible state space,
the perturbation algorithm implements the entangled structure `on top' of the product 
state space. 
An immediate consequence of this approach is that the integrations are carried out 
over the full six-dimensional momentum space of the product representation, instead
of over the five-dimensional parameter space of an irreducible representation.
As a result, we have logarithmically divergent integrals.
It can be easily shown that reducing the number of integration variables by one turns 
the logarithmically divergent integrals of QED into finite integrals.

For practical calculations, the perturbation approach of QED remains the method 
of choice.
This should not be taken as a license to overload the perturbation algorithm with
questionable `physical' interpretations.
There are no indications, neither theoretical nor experimental, that `virtual 
particles' are anything other than elements of a pictorial description of 
momentum entangled structures.
This applies to virtual gauge bosons as well as to virtual electron--positron 
pairs, which, within the perturbation algorithm, are `created' by gauge bosons.

Independently of the partial failure of the perturbation algorithm, it can be
stated that the Poincar\'e group not only determines the structure of the 
elec\-tro\-magnetic interaction, it also explains, by means of the Dynamical Law, 
why there is an electro\-magnetic interaction, and it determines, in contrast to 
the principle of gauge invariance, the correct value of the electromagnetic 
coupling constant.

\subsection{Gravitation}
\label{section_8.2}

This subsection focuses on momentum entangle\-ment in the limit of large 
quantum numbers.
In contrast to the commutation relations between position and momentum, 
the commutation relations between momentum and angular momentum (\ref{2-1}) do not 
contain Planck's constant~$h$. Therefore, they remain valid when $h \rightarrow 0$.
This indicates that the Dynamical Law, derived from these commutation relations, 
also applies to the classical limit.

According to Newton's laws, `exchange of momentum' is another wording for
`curved trajectories'.
From General Relativity it is known that curved trajectories can be described by an 
intrinsic curvature of space--time, such that the trajectories become geodesics 
with respect to the metric of space--time.
Differential geometry describes this curvature by the Riemann 
curvature tensor $R_{\lambda\mu\nu\kappa}$.
The Riemann tensor can be written as 
\begin{eqnarray}
R_{\lambda\mu\nu\kappa} &=&  C_{\lambda\mu\nu\kappa} 
- \frac{1}{6} R \left[ g_{\lambda\nu} g_{\mu\kappa} 
- g_{\lambda\kappa} g_{\mu\nu}\right] \nonumber \\
&+& \frac{1}{2} \left[ g_{\lambda\nu} R_{\mu\kappa} 
- g_{\lambda\kappa} R_{\mu\nu} - g_{\mu\nu} R_{\lambda\kappa}
+ g_{\mu\kappa} R_{\lambda\nu} \right],         \label{9-1}
\end{eqnarray}
where $C_{\lambda\mu\nu\kappa}$ is the traceless Weyl tensor, 
$R_{\mu\nu} = {R^\lambda}_{\mu\lambda\nu}$ is the Ricci tensor,
and $R = {R^\alpha}_\alpha$ is the Ricci scalar \cite{pdm1}.
 
In General Relativity (Einstein--Newton gravity), it is postulated that a certain 
combination of the Ricci tensor and the Ricci scalar be proportional to the 
energy--momentum tensor.
In contrast, the curvature that results from the Dynamic Law is based 
on the exchange or flux of momentum between the particles. 
Unlike Einstein--Newton gravity, this is not a postulate, but rather a direct 
consequence of the momentum entangled structure of two-particle states.
The flux of momentum associated with a particle at a point $x$ is described by 
the off-diagonal elements $T^{0i}(x)$ and $T^{ik}(x)$ of the energy--momentum 
tensor $T^{\mu\nu}(x)$.
There is no direct contribution from the diagonal elements $T^{ii}(x)$.
Therefore, the curvature discussed here is generated by the traceless part of the 
energy--momentum tensor.
In consequence, the generated curvature tensor must be traceless too, which 
means that the curvature must be described by the traceless Weyl tensor.

As shown in \cite{pdm1}, this requirement lead to the field equations
\begin{eqnarray}
W^{\mu\nu} = \alpha_g \, \left(T^{\mu\nu}
 - \frac{1}{2}\,g^{\mu\nu}\,{T^\lambda}_\lambda \right). \label{9-4}
\end{eqnarray}
The tensor on the right is the energy--momentum tensor minus its trace.
The tensor $W^{\mu\nu}$ on the left has the form 
\begin{eqnarray}
W^{\mu\nu} &=& \frac{1}{2} g^{\mu\nu} {({R^\alpha}_\alpha)^{;\beta}}_{;\beta}
+ {R^{\mu\nu;\beta}}_\beta + {R^{\mu\beta;\nu}}_\beta {R^{\mu\beta;\mu}}_\beta 
\nonumber \\
&-& 2 R^{\mu\beta} {R^\nu}_\beta  
+ \frac{1}{2} g^{\mu\nu} R_{\alpha\beta} R^{\alpha\beta}
- \frac{2}{3} g^{\mu\nu} {({R^\alpha}_\alpha)^{;\beta}}_{;\beta} 
\nonumber \\
&+& \frac{2}{3} ({R^\alpha}_\alpha)^{;\mu;\nu} 
+ \frac{2}{3} {R^\alpha}_\alpha R^{\mu\nu} 
- \frac{1}{6} g^{\mu\nu} ({R^\alpha}_\alpha)^2 .  \label{9-5}
\end{eqnarray}
The constant $\alpha_g$ acts as a coupling constant.

Equations (\ref{9-4}) are nothing other than the field equations of conformal 
gravity.
Conformal gravity is known to yield the same results as Einstein--Newton 
gravity as long as systems with the symmetry of the solar system are considered
\cite{pdm2}, see also \cite{jm}.
On the scales of galaxies, the theories differ insofar as conformal gravity is
able to describe the kinematics of galaxies solely on the basis of the distribution of 
visible matter \cite{pdm2}, 
whereas in Einstein--Newton gravity large amounts of properly placed `dark matter' 
are required to match the observational data.
 
What can be said about the value of the coupling constant $\alpha_g$?
A first calculation for a two-particle system of spinless particles, in analogy
to the calculation of the electromagnetic coupling constant, reproduces
this very same coupling constant.
However, in contrast to the electromagnetic interaction, which can be neutralized, 
because there are positive and negative char\-ges, the `gravitational charge'
always has the same sign, which makes it impossible to shield gravity.
Therefore, to determine an effective coupling constant, in principle all 
particles of the universe must be taken into account.
The number of protons derived from an estimated total mass of the visible universe
of $10^{53}$ kg (cf., e.g., \cite{dv} and references cited therein) is $10^{80}$.
To exchange momentum with another particle, a given particle must, at 
first, find a second particle, before it can form a two-particle state with 
this partner.
There are $10^{80}$ different potential partners.
The quantum mechanical transition amplitude for a given particle to form a 
two-particle state with a specific second particle is therefore given by 
$10^{-40}$, the square root of $10^{-80}$.
Hence the strength (or, better, the weakness) of the gravitational interaction is, 
first of all, determined by the factor $10^{-40}$. 
This factor agrees quite well with the empirical ratio between the strengths 
of the gravitational and the electromagnetic interactions.
It has no relation to the Planck mass, but rather to the number of particles in 
the universe.

Although the field equations (\ref{9-4}) define a classical theory of 
gravitation, they describe, in the limit of large quantum numbers, nothing 
other than a basic property of relativistic quantum mechanics, namely the 
momentum entanglement of two-particle states.
(Note that in momentum entangled two-particle states, there are none of those
`ghosts' that in the past have been associated with conformal gravity.)
Conformal gravity is therefore compatible with relativistic quantum mechanics 
and especially with the Dynamical Law.

\section{Conclusions}
\label{section_9}

The Dynamical Law, which requires two-particle states to be momentum entangled, 
is mathematically derivable from the structure of the Poincar\'e group in combination 
with the basic principles of quantum mechanics.
Therefore, as long as neither Poincar\'e invariance nor quantum mechanics is called
into question, this law must be taken into account on the same footing as the 
conservation laws of energy, momentum, and angular momentum.

The Dynamical Law explains the success of gauge theories and at the same time it
helps to understand their deficiencies.
It is more restrictive than the principle of gauge invariance since it
determines the coupling constants.
On the other hand, it covers a wider range than that principle, since it does not 
require the gauge field to be massless.
Therefore, in applying this law to the weak interaction, there is no need for the 
Higgs mechanism, but such a mechanism may well be needed in order to understand the 
different masses of leptons.

There is strong evidence that the interaction mechanism determined by the Dynamical 
Law is responsible for both the electromagnetic and the gravitational interaction. 
Therefore, quantum mechanics and gravitation can no longer be seen as incompatible.
The analysis of how and to what extent quantum electrodynamics and Einstein--Newton 
gravity comply with the Dynamical Law identifies deficiencies in both theories, 
deficiencies that apparently contribute to the well-known problems of the divergences 
in QED and the failure of Einstein's general relativity to describe the kinematics of 
galaxies solely on the basis of visible matter.

The Standard Model describes two more interactions: the strong interaction and the 
weak interaction, which, similarly to QED, are described by the exchange of virtual 
gauge particles and result in momentum entangled structures. 
In quantum chromodynamics, momentum entanglement exists between the constituents of
compound particles, whereas in the weak interaction the exchanged virtual quanta have 
a mass.
In order to trace back these interactions explicitly to a conserved angular momentum, 
a more detailed theoretical understanding of the constituents of baryonic particles on 
the one hand and of the mechanism that is responsible for the different lepton masses 
on the other hand seems to be required.

\appendix
\section*{Appendix: Calculation of the coupling constant}
\setcounter{section}{1}

The following outlines the calculation of the factor $\omega^2$.
A more detailed description can be found in \cite{ws2}.

First consider the 5-dimensional unit ball $B^5$, given by
\begin{equation}
x_1^2 + x_2^2 + x_3^2 + x_4^2 + x_5^2 \le 1.                             \label{a-0}
\end{equation}
Its boundary $\partial B^5$ is the 4-dimensional unit sphere $S^4$
\begin{equation}
x_1^2 + x_2^2 + x_3^2 + x_4^2 + x_5^2 = 1.                             \label{a-1}
\end{equation}
The volume of $B^5$ can be expressed by the integral
\begin{equation}
V(B^5) = \int_0^1\!dr \; r^4 \! \int_{\partial B^5}\!dx_1\,dx_2\,dx_3\,dx_4.    \label{a-4}
\end{equation}
This integral will be used to study the conversion of a spherical volume element
into an Cartesian one (cf.\,Section 5, replacement (\ref{5-3})).
At a given point $b$ on $\partial B^5$, the infinitesimal spherical volume element 
$dr\,dx_1\,dx_2\,dx_3\,dx_4$																			       
\noindent has a rectangular form with four edges of the same (infinitesimal) length
and a fifth, along $r$, with a different length.
Now replace the second integral in (\ref{a-4}) by a 4-dimensional cube in the tangent 
space at $b$, such that the value of the integral is unchanged (`quadrature of the circle').
Then each edge of the cube will have length $s = V(B^5)^{\frac{1}{4}}$.
A 5-dimensional cuboid with volume $V(B^5)$ can be formed in combination with the 
interval $[0,1]$ on the $r$-axis.
By subdividing this cuboid into $N$ intervals in each direction, one obtains a cover of 
the cuboid by $N^5$ volume elements.
Multiplying the length of the edges in the radial direction by $s$ and letting 
$N \rightarrow \infty$ results in a cubical (Cartesian) infinitesimal volume element $d^5\!x$.
The replacement of $dr$ by $s\,dx_5$ introduces a factor of $V(B^5)^{\frac{1}{4}}$.

As mentioned in Section 5, a two-particle state, represented by an integral over product 
states, requires a normalization factor related to the volume of the integration area. 
This volume is defined by the spatial angle covered by the integral (\ref{5-2}).
The normalization factor is contained implicitly in $d\omega(\mathbf{p}_1,\mathbf{p}_2)$ 
but is made explicit in $\omega \, d^5\!p$ (cf.\,replacement (\ref{5-3})).
By taking into account such normalization, the factor $V(\partial B^5)^{-1}$ is introduced 
into $d^5\!x$. 
 
To complicate matters, restrict $\partial B^5$ by the two equations
\begin{equation}
x_1^2 + x_2^2 + x_3^2 = r^2  \;\;\mbox{ and }\;\; x_4^2 + x_5^2 = 1-r^2, \;\; r \le 1.   
\label{a-7}
\end{equation}
Equation (\ref{a-1}) still holds, but whereas $\partial B^5$ has five orthogonal rotational
symmetries, the space (\ref{a-7}) is determined by rotations around $3 + 1$ orthogonal 
rotational axes.
The addition of only one symmetry axis, for example, the rotational axis perpendicular 
to the $x_3$--$x_4$ plane, would convert the restricted space (\ref{a-7}) into (\ref{a-1}).
The integral (\ref{a-4}), calculated for the restricted space, is therefore smaller than 
the original one by a factor determined from the volume of the homogeneous space 
SO(5)/SO(4) $\cong S^4$.
If, however, the restriction (\ref{a-7}) is ignored and $d^5\!x$ is maintained as a 
5-dimensional volume element, for example because the integral is part of an algorithm 
that is based on a 5-dimensional Euclidean parameter space, then at least the result needs 
to be corrected (`renormalized') by the factor $V(S^4)^{-1}$.

These three factors combine to give the expression
\begin{equation}
V(B^5)^{\frac{1}{4}} \, / \, (V(\partial B^5) \, V(S^4)),        \label{a-8}
\end{equation}
which, up to the factor $8\pi$, exhibits the basic structure of Wyler's formula (\ref{7-1}).
(Although the derivation of (\ref{a-8}) has been based on the unit ball, the result does 
not depend on the radius of the ball: 
A scaling of the radius is covered by the scaling property of $d^5\!x$.)

The `formula' (\ref{a-8}) can now be applied to the parameter space of two particles 
with equal mass $m$ that are described by an irreducible two-particle representation of 
the Poincar\'e group.
The total momentum $p = p_1 + p_2$ ($p_1$ and $p_2$ are 4-vectors)                  
satisfies the mass-shell relation 
\begin{equation}
p^2 = M^2 .                                              \label{a-23}
\end{equation}
For the relative momentum $q = p_1 - p_2$, one has
\begin{equation}
q^2 = 4m^2 - M^2 \le 0.     \label{a-24}
\end{equation}
Furthermore, the relations
\begin{equation}
2p_1p_2 = M^2 - 2m^2 \;\;\mbox{ and }\;\; 2p_1\,p = 2p_2\,p = M^2  \label{a-25}
\end{equation} 
let the particle momenta maintain a certain angle to the total momentum.
This restricts the 4-vector $q$ to a spacelike plane, which reduces $q$ to a 2-component 
vector in this plane, respectively, to a 2+1-component vector in space--time.

Equations (\ref{a-23}) and (\ref{a-24}) can be combined to
\begin{equation}
p^2 + q^2 = 4 m^2 , \label{a-27} 
\end{equation}
defining a parameter space with an internal structure comparable to the space defined by
(\ref{a-7}). 
It can be embedded into the 5+2-dimensional homogeneous space SO(5,2)\,/\,SO(5)$\times$SO(2), 
similar to the embedding of space (\ref{a-7}) into space (\ref{a-1}).
By replacing in (\ref{a-8}) the volumes $V(B^5)$ and $V(\partial B^5)$ by the 
corresponding volumes $V(D^5)$ and $V(Q^5)$ , where 
\begin{equation}
V(D^5) = \frac{\pi^5}{2^4\, 5!}, \;\;     	
V(Q^5) = \frac{8 \pi^3}{3}, \;\;          	
V(S^4) = \frac{8 \pi^2}{3} ,        	  \label{a-28}
\end{equation}
Wyler's formula (\ref{7-1}), up to the factor of $8\pi$, is obtained.
The volumes of $D^5$ and $Q^5$ have been calculated in \cite{hua} from a bounded 
realization of SO(5,2)\,/\,SO(5)$\times$SO(2) on the complex unit ball.

Two more factors, leading to the contribution of $8\pi$ in Wyler's formula, still have to be 
added:
Firstly, a factor of $2$ as part of the Jacobian that relates $p$ and $q$ to $p_1$ and $p_2$.
Secondly, a factor of $4\pi$, because in the momentum representation (the $p$-representation),
the effective coupling constant in QED, to which must be compared the factor $\omega^2$ as 
part of the transition amplitude (\ref{6-1}), is not $e$ but $e / 4\pi$.
Together with the convention $\alpha = e^2 / 4\pi$, this results in the factor $4\pi$.

The normalization factor $\omega$ of a two-particle state is, as usual, obtained as the square
root of the volume factor $\omega^2$.

\end{document}